\documentclass[twocolumn]{aastex63}

\newcounter{daggerfootnote}

\newcommand{\jwst}{\textit{JWST}}
\newcommand{\hst}{\textit{HST}}

\definecolor{mycol}{rgb}{0,0,1}

\received{n/a}
\revised{n/a}
\accepted{n/a}

\submitjournal{\apj}

\shorttitle{Photometric Census of Red Dots}
\shortauthors{Kokorev et al.}

\usepackage{xcolor}
\usepackage{longtable}
\usepackage{lipsum}
\usepackage{amsmath}
\usepackage[normalem]{ulem}
\usepackage[para]{threeparttable}
\usepackage{enumitem}
\usepackage[encapsulated]{CJK}

\begin{document}

\title{A Census of Photometrically Selected Little Red Dots at $4<z<9$ in JWST Blank Fields}

\correspondingauthor{Vasily Kokorev}
\email{kokorev@astro.rug.nl}

\author[0000-0002-5588-9156]{Vasily Kokorev}
\affiliation{Kapteyn Astronomical Institute, University of Groningen, 9700 AV Groningen, The Netherlands}

\author[0000-0001-8183-1460]{Karina I. Caputi}
\affiliation{Kapteyn Astronomical Institute, University of Groningen, 9700 AV Groningen, The Netherlands}
\affiliation{Cosmic Dawn Center (DAWN), Niels Bohr Institute, University of Copenhagen, Jagtvej 128, K{\o}benhavn N, DK-2200, Denmark}

\author[0000-0002-5612-3427]{Jenny E. Greene}
\affiliation{Department of Astrophysical Sciences, Princeton University, 4 Ivy Lane, Princeton, NJ 08544}

\author[0000-0001-8460-1564]{Pratika Dayal}
\affiliation{Kapteyn Astronomical Institute, University of Groningen, 9700 AV Groningen, The Netherlands}

\author[0000-0002-6849-5375]{Maxime Trebitsch}
\affiliation{Kapteyn Astronomical Institute, University of Groningen, 9700 AV Groningen, The Netherlands}

\author[0000-0002-7031-2865]{Sam E. Cutler}
\affiliation{Department of Astronomy, University of Massachusetts, Amherst, MA 01003, USA}

\author[0000-0001-7201-5066]{Seiji Fujimoto}
\affiliation{Department of Astronomy, The University of Texas at Austin, Austin, TX 78712, USA}
\affiliation{Cosmic Dawn Center (DAWN), Niels Bohr Institute, University of Copenhagen, Jagtvej 128, K{\o}benhavn N, DK-2200, Denmark}

\author[0000-0002-2057-5376]{Ivo Labb\'e}
\affiliation{Centre for Astrophysics and Supercomputing, Swinburne University of Technology, Melbourne, VIC 3122, Australia}

\author[0000-0001-8367-6265]{Tim B. Miller}
\affiliation{Center for Interdisciplinary Exploration and Research in Astrophysics (CIERA), Northwestern University, 1800 Sherman Ave, Evanston IL 60201, USA}

\author[0000-0001-8386-3546]{Edoardo Iani}
\affiliation{Kapteyn Astronomical Institute, University of Groningen, 9700 AV Groningen, The Netherlands}

\author[0000-0001-6066-4624]{Rafael Navarro-Carrera}
\affiliation{Kapteyn Astronomical Institute, University of Groningen, 9700 AV Groningen, The Netherlands}

\author[0000-0002-5104-8245]{Pierluigi Rinaldi}
\affiliation{Kapteyn Astronomical Institute, University of Groningen, 9700 AV Groningen, The Netherlands}

\begin{abstract}
Observations with the James Webb Space Telescope (JWST) have uncovered numerous faint active galactic nuclei (AGN) at $z\sim5$ and beyond. These objects are key to our understanding of the formation of supermassive black holes (SMBHs), their co-evolution with host galaxies, as well as the role of AGN in cosmic reionization. Using photometric colors and size measurements,  we perform a search for compact red objects in an array of blank deep JWST/NIRCam fields totaling $\sim640$ arcmin$^{2}$. Our careful selection yields 260 reddened AGN candidates at $4<z_{\rm phot}<9$, dominated by a point-source like central component ($\langle r_{\rm eff} \rangle <130$ pc) and displaying a dichotomy in their rest-frame colors (blue UV and red optical slopes). Quasar model fitting reveals our objects to be moderately dust extincted ($A_{\rm V}\sim1.6$), which is reflected in their inferred bolometric luminosities of $L_{\rm bol}$ = 10$^{44-47}$ erg/s, and fainter UV magnitudes $M_{\rm UV} \simeq$ $-17$ to $-22$. Thanks to the large areas explored, we extend the existing dusty AGN luminosity functions to both fainter and brighter magnitudes, estimating their number densities to be $\times100$ higher than for UV-selected quasars of similar magnitudes. At the same time 
they constitute only a small fraction of all UV-selected galaxies at similar redshifts, but this percentage rises to $\sim$10\% for $M_{UV}\sim -22$ at $z\sim7$. Finally, assuming a conservative case of accretion at the Eddington rate, we place a lower limit on the SMBH mass function at $z\sim5$, finding it to be consistent with both theory and previous JWST observations.
\end{abstract}

\keywords{Active galactic nuclei (16), High-redshift galaxies (734), Early universe (435)}

\section{Introduction} \label{sec:intro}
The remarkable sensitivity and angular resolution of the James Webb Space Telescope (\textit{JWST}) at infrared wavelengths is enabling us to explore the distant Universe like never before. This allows for an exceptionally detailed examination of the characteristics of known high-$z$ sources \citep[e.g.][]{bunker23, maiolino23} and, at the same time, reveals the presence of more and farther galaxies \citep[e.g.][]{adams23,atek23a, austin23,bradley23,casey23,finkelstein23,naidu22, robertson23}, some of them spectroscopically confirmed beyond $z>13$ \citep{curtislake23,wang23}.
\par
What truly tests our models and preconceived vision of galaxy evolution is not how early we can see these objects, but the questions they raise regarding the balance between their mass, UV luminosity and age.
The excess of high-$z$ galaxies at the bright end ($M_{\rm UV}\leq-20$) of the UV luminosity function
is in tension with current theoretical frameworks \citep{behroozi15,dayal17,yung19,yung20,behroozi19,behroozi20,dave19,wilkins22,kannan23,mason23,mauerhofer23}, which suggests exotic initial mass functions, little to no dust attenuation, or a higher than 
anticipated density of galaxies undergoing active galactic nuclei (AGN) phenomena \citep[e.g.][]{finkelstein22,pacucci22, boylan-kolchin23,ferrara23,fujimoto23_uncover,lovell23,steinhardt23, sun_g23}.
\par
Although early hints also existed in prior works \citep{morishita20,fujimoto22,endsley23}, 
one of the most intriguing discoveries from early \textit{JWST} imaging is that of compact red sources with a ``v-shaped'' spectral energy distribution (SED), namely a blue UV continuum and a steep red slope in the rest-frame optical \citep{labbe23_nat,labbe23,furtak23_phot}. While the first photometric selections of these objects included spatially resolved targets that could be early massive compact galaxies \citep{barro23}, spectra revealed clear evidence for broad H$\alpha$ and/or H$\beta$ emission indicative of actively accreting supermassive black holes \citep[SMBH;][]{furtak23_nat,fujimoto23_uncover,greene23,killi23,kocevski23,kokorev23c,matthee23,ubler23}.
\par
 Dubbed ``little red dots'' (LRDs), these sources have SEDs characterized by a unique ``v-shaped'' continuum combined with their point source morphology \citep{labbe23_nat,labbe23,furtak23_phot}. However, what truly makes the LRDs stand out is their high number densities. It appears that LRDs may account for a few percent of the galaxy population at $z>5$, and are far more numerous than the lowest luminosity known UV-selected quasars. Likewise, they appear to account for $\sim 20\%$ of broad-line selected active galactic nuclei (AGN) at $z \sim 5-6$ \citep{greene23,harikane23_agn,labbe23,maiolino23b}, which is higher than the fraction of dusty red quasars at $z<2$ 
 \citep{banerji15,glikman15}. These red dots are generally observed at $z\sim5$ \citep{labbe23}, but can potentially be found even at $z>9$ \citep{leung23}. However, these initial LRD studies were performed with limited spectroscopic samples and/or small areas of the sky, covering only $\sim 20$ -- 40 arcmin$^{2}$. The numbers of compact red objects could therefore be further affected by cosmic variance, which makes it quite difficult to assess their real importance and diversity.



Extending the selection of this compact red population of low-luminosity broad-line AGN candidates to larger areas would thus be necessary to study their complete demographics, limiting the effects of cosmic variance. In addition, this would provide us with a sufficient level of detail toward a better understanding of the total number densities of obscured AGN at high-$z$ as well as the potential role that these sources play in cosmic reionization \citep[e.g. see][]{grazian18,mitra18,dayal20,trebitsch23,dayal24}.
\par
In this work we present a carefully selected sample of 260 reddened AGN candidates in the $\sim640$ arcmin$^{2}$ area covering some of the deepest blank extragalactic \jwst\, fields. Examining such a large area will ensure that we are reducing the effects of cosmic variance to a minimum, while our focus on blank fields lessens the selection biases and avoids volume uncertainties arising from lensing magnification.
\par
Throughout this work we assume a flat $\Lambda$CDM cosmology \citep[e.g.][]{planck20} with $\Omega_{\mathrm{m},0}=0.3$, $\Omega_{\mathrm{\Lambda},0}=0.7$ and H$_0=70$ km s$^{-1}$ Mpc$^{-1}$, and a \citet{chabrier} initial mass function (IMF) between $0.1-100$ $M_{\odot}$. All magnitudes are expressed in the AB system \citep{oke74}.

\begin{figure*}
\begin{center}
\includegraphics[width=.9\textwidth]{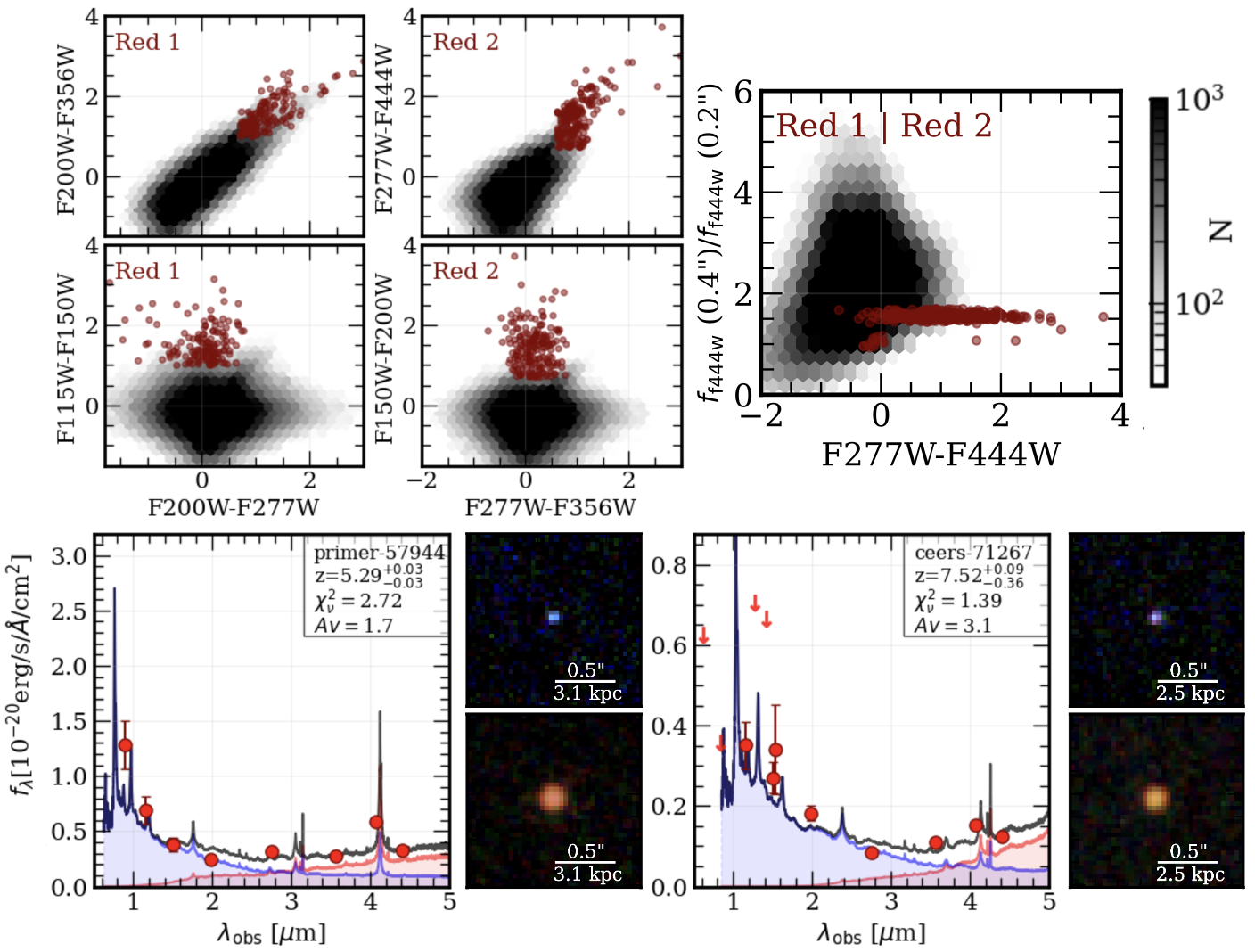}
\caption{Selection and analysis of LRD candidates. \textbf{Top:} Sample selection criteria. The left and central panels show modified ``\texttt{red 1}'' ($z\lesssim 6$) and ``\texttt{red 2}'' ($z\gtrsim 6$) color-color cuts from \citet{labbe23}.  The right panel shows the compactness cut of our sample. Selected objects are highlighted as maroon circles, while grayscale hexbins show the full catalog. The compact red sources are clear outliers in color-color-compactness space. Colorbar is shared between all plots. \textbf{Bottom:} An example of best-fit SEDs to the photometry of LRD candidates with the dust-free (blue) and dusty (red) AGN templates \citep{vandenberk01,glikman06} at representative redshifts of $z\sim6$ and $z\sim8$. The combined model is shown in black.  Detections ($>3\sigma$) are shown as red circles, while upper limits (primarily from \textit{HST}) are shown as downward arrows. On the right of each SED we show 1\farcs{5} color composite cutouts in the short (F115W/F150W/F200W) and long (F277W/F356W/F444W) NIRCam filters.}
\label{fig:fig1}
\end{center}
\end{figure*}

\section{Observations and Data} \label{sec:obs_data}
\label{sec:data}

In this work we use \textit{JWST} data from the following programs/fields - CEERS (\# 1345, PI: S. Finkelstein, \citealt{bagley22}) in the EGS, PRIMER (\# 1837, PI: J. Dunlop) in COSMOS and UDS. For the GOODS-S we combine the available data from multiple broad and medium band programs - FRESCO (\# 1895, PI: P. Oesch, \citealt{oesch23}), JADES (\# 1180, 1210, 1286, 1287 PIs: D. Eisenstein, N. Luetzgendorf, \citealt{eisenstein23,eisenstein23b}) and JEMS (\# 1963, PI: C. Williams, \citealt{williams23a}). We provide a general overview of these 4 fields in \autoref{tab:tab1}. More detailed information, including specific filters, depths and survey designs can be found in overview papers for each data release.

\subsection{$JWST$ Imaging Data Reduction}
We homogeneously processed all the publicly available \jwst\ imaging obtained with the NIRCam and MIRI in a variety of public \textit{JWST} fields, presented in \autoref{tab:tab1}.
The images have all been reduced with the \textsc{grizli} pipeline \citep{grizli}, using the \texttt{jwst\_1084.pmap}, and follow the same methodology of (multiple) previous studies \citep[e.g.,][]{jin23,kokorev23a,valentino23}. Compared to the standard pipeline, we incorporate additional corrections to account for cosmic rays and stray light \citep[see e.g.,][]{bradley23}, 1/f noise, detector level artifacts (``wisps'' and ``snowballs'') and bias in individual exposures \citep[see e.g.,][]{rigby23}. For the PRIMER data, we introduce an additional procedure that alleviates the detrimental effects of the diagonal striping seen in some exposures as was done in \citet{valentino23}. Finally, our mosaics include the updated sky flats for all NIRCam filters. These reductions are publicly available as a part of the DAWN \jwst\, Archive (DJA\footnote{\url{dawn-cph.github.io/dja/}}).
\par 
These data-sets are further complemented by including all available optical and near-infrared data from the Complete \textit{Hubble} Archive for Galaxy Evolution \citep[CHArGE,][]{kokorev22}. Individual \jwst\ and \hst\ exposures were aligned to the same astrometric reference frame by using the Gaia DR3 \citep{gaia-collaboration2021}, then co-added and drizzled \citep{fruchter2002} to a $0.\arcsec04$ pixel scale for all the  \jwst\ and \hst\ filters.
\par
Some of the fields we examine in this work have also been observed with MIRI, in one or more filters, sampling mostly the rest-frame near-infrared (NIR) at $z\gtrsim4$. These data are, however, not uniform in the  wavelength coverage, depth and area. In fact, only about a third of the objects in areas we examine have public MIRI data and even fewer are actually detected. While the inclusion of the MIRI photometry can assist in further identifying the presence (or absence) of dusty, power law-like AGN component in galaxies \citep[e.g. see][]{yang23,williams23b}, doing so appropriately within a context of
a population study requires a degree of uniformity which the current MIRI data do not possess. Therefore, we have opted to exclude MIRI photometry from our current analysis to maintain consistency across various fields.

\subsection{Source Extraction}
The initial \textit{JWST} catalog was constructed by utilizing a detection image combined from all noise weighted ``wide'' (W) NIRCam Long Wavelength (LW) filters available, which includes F277W, F356W and F444W. A similar detection method was already successfully employed in several works (see e.g. \citealt{jin22b,kokorev23a,valentino23,weaver23}). To extract the sources and produce a segmentation map, we used \textsc{sep} \citep{sep}, a \textsc{Python} version of \textsc{SExtractor} \citep{sextractor}. Photometry was extracted in circular apertures of increasing size. Correction from the aperture to the ``total'' values was performed by using the \texttt{flux\_auto} column output by \textsc{sep}, which is equivalent to the \texttt{MAG\_AUTO} from \textsc{SExtractor}, ensuring that for each source only flux belonging to its segment is taken into account. This method was shown to apply to both point-like and extended objects \citep{weaver21,weaver23_farmer}, so we believe it to be adequate for our sources.
\par
Additionally, we introduce a correction to account for the missing flux outside the Kron aperture \citep{kron1980}, by utilizing a method similar to the one used in \citet{whitaker11} and  \citet{weaver23}. In short, this procedure involves computing the fraction of the missing light outside the circularized Kron radius by analyzing curves of growth of the point spread functions (PSF), which were obtained empirically, by stacking stars in these various fields.  This correction is then applied to the \texttt{flux\_auto} values for each source. However, since our work focuses on compact (sub NIRCam PSF size) AGN candidates this additional correction does not strongly influence the derived flux densities. For the same reason, we use the total fluxes, computed from $D$=0\farcs{36} apertures, unless specified otherwise.


\begin{deluxetable*}{ccccc}
    \tablecaption{Properties of the observed fields with \jwst/NIRCam observations.\label{tab:tab1}}
    \tablehead{
    \colhead{Field}&
    \colhead{R.A. [deg]}&
    \colhead{Dec. [deg]}&
    \colhead{Science Area [arcmin$^2$] }&
    \colhead{NIRCam depths [mag]}}
    \startdata
CEERS & 214.920 & 52.870  & 97.0 & 
            $28.8\,/28.5\,/28.7\,/28.9\,/29.0\,/28.4$  \\ 
PRIMER-COSMOS  & 150.119 &  2.325 &  197.2 & 
            $27.9\,/28.1\,/28.3\,/28.7\,/28.6\,/28.2$ \\ 
PRIMER-UDS & 34.372 & -5.210 & 274.6 & 
            $27.6\,/27.8\,/28.0\,/28.3\,/28.4\,/28.0$  \\
GOODS-S &  53.142 & -27.798 & 69.7 & $29.6\,/29.6\,/29.5\,/29.8\,/29.6\,/29.3$  \\    
\enddata
\tablecomments{\textbf{NIRCam depths:} expressed as $5\sigma$ within the $0\farcs36$ apertures used for the photometric extraction in the area covered by F115W/F150W/F200W/F277W/ F356W/F444W.}
\end{deluxetable*}

\begin{figure}
\begin{center}
\includegraphics[width=.49\textwidth]{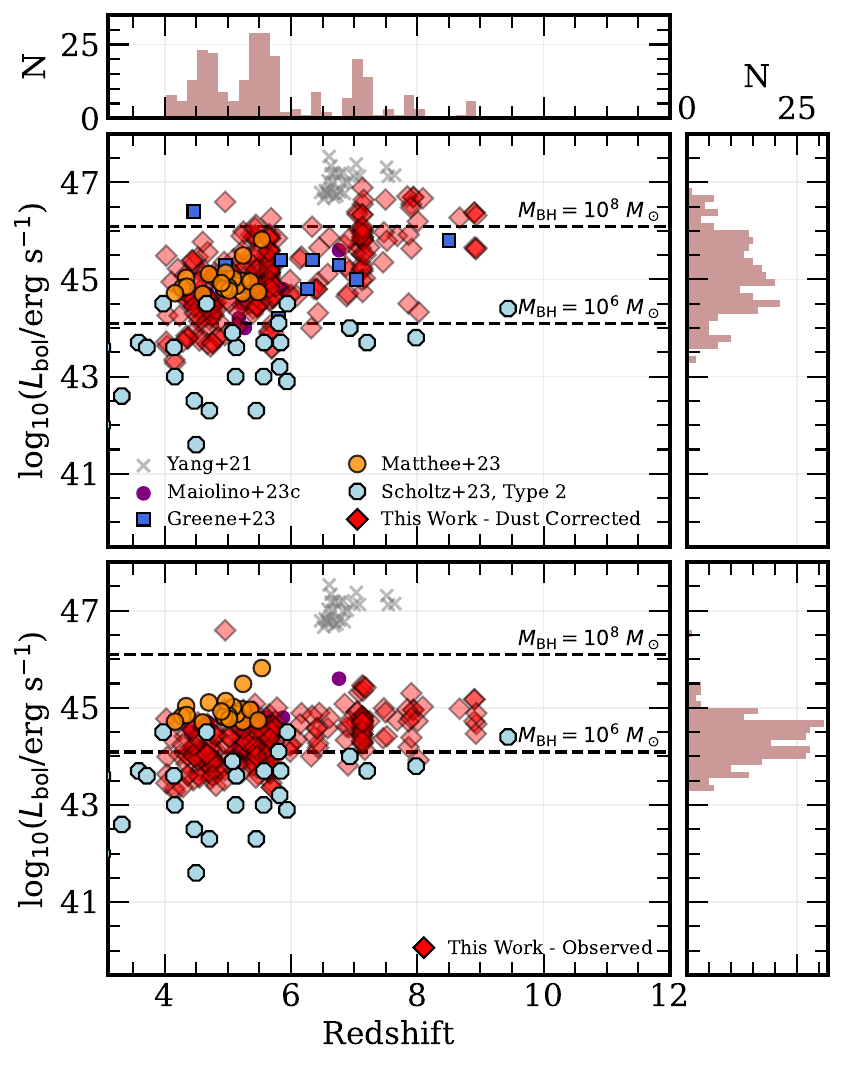}
\caption{\textbf{Top:} Distribution of the $L_{\rm bol}$ assuming an AGN dominated rest-frame optical continuum, with best-fit dust correction applied. Our compact sources span a wide range of luminosities across the redshift range of interest. We contrast our results to the BL AGN from \citet[][blue squares]{greene23}, \citet[][orange circles]{matthee23}, $z>6.5$ quasars from \citet[][gray crosses]{yang21}, high-$z$ AGN from \citet[][magenta circles]{maiolino23b} and finally objects hosting Type 2 AGN from \citet[][light blue circles]{scholtz23}. Assuming $\lambda_{\rm edd}=1$, we show what $L_{\rm bol}$ would correspond to $M_{\rm BH}=10^{6-8} M_\odot$ (dashed lines). \textbf{Bottom:} Same as before, but we show the distribution of $L_{\rm bol}$ without correcting for the best-fit dust-attenuation. Our final sample has a median $A_{\rm V}\sim1.6^{+1.1}_{-1.0}$. On the side of each panel we also show histograms highlighting the redshift and $L_{\rm bol}$ distributions.}
\label{fig:fig2}
\end{center}
\end{figure}

\section{Identifying Compact Red Objects} \label{sec:ident}
The data from CEERS, PRIMER and various programs covering GOODS-S are well-suited for a photometric search for compact obscured AGN candidates.
The available \textit{HST} and \textit{JWST} photometry covers a complete wavelength range from 0.4 -- 5 $\mu$m,  in at least $7$ broad and medium bands, reaching a median 5$\sigma$ depth of 28.3 AB mag in F444W filter (\autoref{tab:tab1}).  In our search, we explore blank fields covering a large area of $\sim 640$ arcmin$^{2}$, which are also completely independent. In return this will significantly limit the impact of cosmic variance and enable us to avoid dealing with the cosmic volume uncertainties introduced by the lensing magnification.

\subsection{Color and Morphology Selection}
Recently, \citet{labbe23} published a large sample of photometrically identified compact red sources from the Cycle 1 \textit{JWST} UNCOVER program \citep[PIs: I. Labb\'e, R. Bezanson;][]{bezanson22}. Subsequent follow-up of 17 such objects with NIRSPec/MSA PRISM has resulted in a remarkable success rate of 83 \%, with 14/17 photometrically selected targets confirmed as broad-line (BL) AGN at $4<z<8.5$ \citep{greene23}, and 3/17 as brown dwarfs \citep{burgasser23}. In brief, the color cuts introduced in \citet{labbe23} are designed to catch the break between the red continuum slope in rest-frame optical, and the blue rest-UV emission ($\lambda_{\rm rest}\sim4000$ \AA\,). This color selection requires that the red continuum slope is rising in more than one adjacent filter pair, to avoid selecting galaxies with strong emission lines. Indeed, currently available spectra of LRDs \citep[e.g.][]{fujimoto23_uncover,furtak23_nat,kocevski23,kokorev23c,greene23,matthee23} display a remarkable dichotomy in their observed spectral shapes. In particular the SEDs at $1-2$ $\mu$m (1000 - 2000 \AA\, rest) are blue ($f_\lambda\propto\lambda^{-2}$) and red ($f_\lambda\propto\lambda^{0-2}$) at $3-5$ $\mu$m (3100 - 5200 \AA\, rest). As such we keep the \citeauthor{labbe23} color criteria largely unchanged, only introducing some further adjustments based on the UNCOVER spectra of LRDs, namely to limit the contamination of our sample by brown dwarfs as was suggested in \citet{greene23}.
\par
Colors alone would end up selecting both LRDs and extended red galaxies \citep[see e.g.][]{labbe23,williams23b,williams23a}, so we introduce a further ``compactness'' cut to only select sources with high central flux concentration. To do that we use the ratio between the total flux in F444W between 0\farcs{4} and 0\farcs{2} apertures. Since roughly 17\% of the LRD candidates followed up with NIRSpec turned out to be brown dwarfs \citep{burgasser23}, we would also like to minimize the incidence of these objects in our sample. To do that we adopt the brown dwarf removal criterion from \citet{greene23}, based on the LRD spectra from NIRSPec/MSA. Finally, we also require our sources to be significantly ($>14 \sigma$) detected in F444W, and be brighter than 27.7 AB mags, to be consistent with the UNCOVER selection. The imposed color cuts are then:
\begin{equation*}
\begin{aligned}
\texttt{red} \, \texttt{1} & = \mathrm{F115W}-\mathrm{F150W}<0.8 \quad \& \\
&\mathrm{F200W}-\mathrm{F277W}>0.7 \quad \& \\
&\mathrm{F200W}-\mathrm{F356W}>1.0  
\end{aligned}
\end{equation*}
or
\begin{equation*}
\begin{aligned}
\texttt{red} \, \texttt{2} & = \mathrm{F150W}-\mathrm{F200W}<0.8 \quad \& \\
&\mathrm{F277W}-\mathrm{F356W}>0.6 \quad \& \\
&\mathrm{F277W}-\mathrm{F444W}>0.7,
\end{aligned}
\end{equation*}
which are effectively selecting our low ($z<6$) and high ($z>6$) redshift samples, respectively. The compactness is given by:
\begin{equation*}
\texttt{compact} = f_{\rm f444w} (0\farcs{4})/f_{\rm f444w} (0\farcs{2})<1.7.
\end{equation*}
To limit the number of brown dwarfs in the sample we also adopt:
\begin{equation*}
\texttt{bd\_removal} = \mathrm{F115W} - \mathrm{F200W}>-0.5.
\end{equation*}
The final selection then becomes ($\texttt{red} \, \texttt{1} \, | \,\texttt{red} \, \texttt{2}$) \& \texttt{compact} \& \texttt{bd\_removal}.
Applying the color criteria also means that every object has to be detected ($>3\sigma$) in at least one band per color to make the selection meaningful. In case of a non-detection we use the $2\sigma$ upper limits, but only if the ``brighter'' band in the color is detected. Out of $\sim$ 408 000 objects covering 4 fields of interest, we end up selecting 334. Most importantly, we note that no information about photometric redshifts and underlying galaxy/AGN SEDs is used at this stage to avoid being biased by models. We discuss our photometric redshift estimate and its agreement with spec-$z$ for sub-samples in the next sub section.

\begin{figure}
\begin{center}
\includegraphics[width=.49\textwidth]{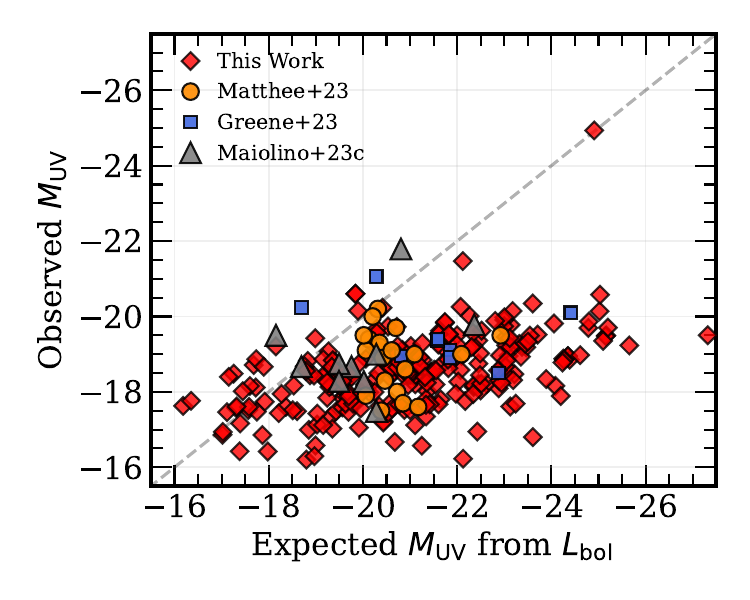}
\caption{Observed $M_{\rm UV}$ compared to the $M_{\rm UV}$ expected from the dust-corrected $L_{\rm bol}$. We derive the expected $M_{\rm UV}$ values by following the relation from \citet{shen20}. We also show the data for red dots from \citet{greene23} (blue squares) and \citet{matthee23} (orange circles), as well as broad-line AGN at $z>4$ from \citet{maiolino23b} (gray triangles). The gray dotted line shows the 1:1 relation.
The observed $M_{\rm UV}$ we derive is fainter than the expected value, with $A_{\rm UV}$ varying from $\sim 0.6 - 4.2$. While extreme, the UV attenuation is more than 5 magnitudes smaller when compared to values expected given our best-fit $A_{\rm V}$. }
\label{fig:fig3}
\end{center}
\end{figure}

\subsection{Size Measurements}
 While the compactness cut alone already successfully manages to select PSF-dominated point sources, we would like to provide a further fine-tuning to provide a fully quantitative rather than qualitative assessment. To do that we fit our sources with \textsc{pysersic} \citep{pasha23} in the F444W band. The primary goal of this is to ensure that the source is dominated by the PSF component in the reddest, least dust obscured, band as was done in \citet{labbe23}. 
 We focus on the F444W band for this analysis, as the galactic origin of the rest-UV can not be ruled out with current photometric (or even spectroscopic) observations. Moreover, if an object is dominated by a single star-forming region, it could appear compact in rest-UV bands, but still be extended in the redder filters, making the F444W band the most physically constraining for our type of study.
\par
Taking the PSF into account is imperative when measuring sizes of unresolved objects. We generate our F444W PSFs empirically for each field by following the methodology described in \citet{Skelton14}, \citet{whitaker19} and \citet{weaver23}. In brief, we identify non-saturated stars in every field by considering objects on the stellar locus, that are brighter than 24 AB mags and extract these candidates in stamps. These stamps are then centered and normalized to unity. The final PSFs are derived by averaging the weighted stamps, and are then normalized to match the enclosed energies of the expected \textit{JWST} calibration levels within 4" diameter apertures\footnote{\url{https://jwst-docs.stsci.edu/jwst-near-infrared-camera/nircam-performance/nircam-point-spread-functions}}. For more detail see the Appendix in \citet{weaver23}.
\par
The light is modeled with a single S\'ersic \citep{sersic} profile with the center, brightness, effective radius, S\'ersic index, and axis ratio as free parameters. The prior for the index is uniform between $0.65-6$ and the effective radius uniform between $0.25  - 5$ pixels (0\farcs{01} -- 0\farcs{2} ). For each source we create a 3" square cutout (75 pixels by 75 pixels) and mask any additional sources within the stamp. Parameter values and uncertainties are calculated using the Laplace approximation, assuming that the posterior is Gaussian. We exclude fits where the resulting $\chi^2$ per pixel is greater than 2 or the best fit flux differs from the catalog value by more than 2 AB magnitudes. This excludes 15 sources from our sample which by visual inspection we find are untrustworthy due to contamination of bright nearby objects.

A source can be considered to be point-like if its effective radius in F444W band is lower than the empirical PSF FWHM ($\sim$ 0\farcs{15}). It appears that our \texttt{compact} criterion is extremely effective at identifying PSF-like objects as
none of the 319/334 sources with reliable fits exceed a diameter of 0\farcs{08}, when considering the 95 \% size upper limits, corroborating the effectiveness of the compactness criterion described in \autoref{sec:ident}. After carefully considering both colors and morphology when selecting our sample of AGN candidates, we are now able to proceed directly to the SED fitting.

\subsection{Photometric Redshifts}
\label{sec:ez}
To calculate photometric redshifts ($z_{\rm phot}$) for our objects, we use the \textsc{Python} version of \textsc{EAZY} \citep{brammer08}. We choose the 
\textsc{blue\_sfhz\_13} model subset \footnote{\url{https://github.com/gbrammer/eazy-photoz/tree/master/templates/sfhz}} that contains redshift-dependent SFHs, and dust attenuation values.More specifically, the linear combinations of log-normal SFHs included in the template set are not allowed to exceed redshifts that start earlier than the age of the Universe \citep[for more detail see][]{blanton07}. These models are further complemented by a blue galaxy template, derived from a \textit{JWST} spectrum of a $z=8.50$ galaxy with extreme line equivalent widths \citep[ID4590;][]{carnall22}. 

While it might seem counter-intuitive to use galaxy templates for what we believe to be AGN candidates, similar efforts presented in \citet{labbe23} report a good agreement between deriving $z_{\rm phot}$ with stellar templates alone, as opposed to stellar+AGN models, finding a very good agreement between the two. This is not surprising, as when it comes to photometric redshift fitting, the key deciding factors are the positions of the Lyman ($\sim 912$ \AA) and Balmer ($\sim 4000$ \AA) breaks.

For the LRDs, a general absence of significant stellar contribution in the rest-frame optical \citep[e.g.][]{greene23} would result in a lack of a noticeable Balmer break, however the trough of the ``v-shape'' in the rest-frame SEDs of observed LRDs is also located at roughly 4000 \AA\, \citep[e.g.][]{furtak23_nat, kokorev23c}. Indeed the existence of such a feature in LRDs has resulted in their misidentification as dusty star-forming galaxies, leading to stellar mass estimates which are in tension with $\Lambda$CDM,
if all the light is attributed to star-formation alone \citep[e.g. see discussion in][]{boylan-kolchin23,kocevski23,labbe23_nat,steinhardt23}. 

Spectroscopic follow-up of red compact objects hosting AGN BL emission has in fact shown a remarkable agreement between the $z_{\rm phot}$ derived with \textsc{EAZY} (or similar routines) and $z_{\rm spec}$. For example, in GOODS-S, \citet{matthee23} report an average $\sigma_{\rm z}=|\Delta{\rm z}|/(1+z_{\rm spec})=0.01$, and UNCOVER LRDs presented in \citet{greene23} have shown $\sigma_{\rm z}\sim0.04$. Similar consistency was also found between the initial photometric source selection and final spectra in JADES and CEERS fields \citep{maiolino23b,kocevski23,andika24}. As such we consider that utilizing \textsc{EAZY} to derive redshifts is adequate for our sample.

We fit all the available photometry, and upper limits from HST/F435W ($\lambda_{\rm obs}\sim0.4$ $\mu$m) to JWST/F444W ($\lambda_{\rm obs}\sim4.4$ $\mu$m) filters for our sample of 319 LRDs, limiting the redshift grid between $0.01<z<20$. From the best fit \textsc{EAZY} SEDs we only derive photometric redshifts, delegating the estimation of physical parameters to a different template set discussed in the next section. The uncertainties on the photometric redshift are computed from the 16th and 84th percentiles of the redshift probability distributions - $p(z)$. The availability of \textit{HST} photometry allows us to securely constrain the presence of the Lyman break either through diminishing flux where a given filter overlaps with the break, or via upper limits for 40 \% sources in our sample. Notably, however, the presence of the Lyman break is not required to securely constrain redshift for high-$z$ LRDs, as the break in the optical part of the SED already places these objects in a unique color-color space, as initially shown in \citet{labbe23} and then further confirmed in \citet{greene23}. In addition, access to at least one NIRCam medium band further enhances redshift quality by allowing us to identify emission lines in broad band photometry. As a result none of our objects have double peaked redshift solutions. Despite that, since our sample is still only identified photometrically, appropriately taking into account $z_{\rm phot}$ uncertainties is crucial when deriving the physical parameters and luminosity functions in the upcoming sections.

\subsection{Quasar Template Fitting} \label{subsec:quasar_fit}
While the origin of the rest-frame UV light in LRDs remains elusive, growing samples of \textit{JWST} spectra  consistently show either a complete absence or a lack of a significant contribution from the host galaxy to the total flux in the rest-frame optical ($\lambda_{\rm obs}\gtrsim 2$ $\mu$m) \citep{furtak23_nat,greene23,kokorev23c}. This is generally evidenced by comparing the expected $L_{\rm 5100}$ from broad Balmer series lines (generally H$\beta$ and/or H$\alpha$) to the observed values. For example \citet{greene23} find that H$\alpha$-derived and observed $L_{\rm 5100}$ agree within a factor of two for the objects which have H$\alpha$ PRISM coverage. Supporting this, \citet{furtak23_nat} and \citet{kokorev23c} also find that black holes masses ($M_{\rm BH}$) derived via broad H$\beta$ line luminosity and continuum are identical, given the scatter of the relations derived from AGN reverberation mapping \citep[see e.g.][]{kaspi00,greene05}, hinting at negligible stellar components. Furthermore none of the currently known spectroscopically confirmed LRDs in Abell 2744 are detected in ALMA at 1.2 mm down to $<70$ $\mu$Jy (2$\sigma$), which strongly limits the contribution of obscured star formation \citep[see e.g.][]{labbe13,labbe23}, unless the dust is either very cold, very hot or diffuse. Indeed, when both \textit{JWST} data and ALMA upper limits \citep{fujimoto23,fujimoto23_alma_uncover} are considered in a joint AGN+galaxy template fitting for the objects described in \citeauthor{furtak23_nat} and \citeauthor{kokorev23c} the contribution of the galaxy model to the total rest-frame optical light is negligible. Finally, robust measurements of effective radii for all UNCOVER LRDs, while also taking into account the empirically derived PSFs \citep[see][]{weaver23}, find no strong evidence for extended emission associated with the host galaxy in the F444W band.

Unfortunately, a lack of deep and uniform ALMA coverage for our objects prevents us from carrying out joint AGN+galaxy template fitting to ascertain the amount of AGN contribution to the optical SED. While it is possible to do it with only \textit{JWST} photometry, such a fit would be too degenerate given the available number of bands and the number of models required. 
However, objects in our work were specifically selected with the color and compactness criteria largely mirroring those used to identify broad-line AGN in UNCOVER. It is reasonable therefore to assume that given similarly red ($f_\lambda\propto\lambda^{0-2}$ at 3100 -- 5200 \AA\, rest) slopes, the rest optical continuum in our sources is also dominated by AGN light. 

In terms of luminosity, the dust-obscured component is dominating the light from LRDs, and must be substantially attenuated ($A_{\rm V}\sim1-2$) in order to fit the observed red slope. Given that, the rest-UV light should not be visible at all ($A_{\rm UV}>10$). From our photometry, however, we see that while the blue component is weak (only a few percent of red component), it is not reddened.  This emission can be interpreted as either scattered light from the AGN itself, or the host galaxy \citep[see discussion in][]{labbe23,greene23}. However, even when spectra are available \citep{greene23}, given the similarities between the UV slopes of quasars and young star-forming galaxies, these two models are equally good representations of the observed light. Our available data also do not allow us to make a clear distinction between these two possibilities, therefore to avoid over-interpreting the origins of the rest-UV emission, we will assume the scattered light (unreddened) only template in our modeling. We caution the reader that as a result of the unknown origin of the blue light, the rest-UV properties derived in this paper do not necessarily represent physical conditions of the potential AGN our LRDs might host. Due to the aforementioned similarity between UV slopes in quasars and SFGs, the $M_{\rm UV}$ values derived from both galaxy and quasar fits are thus nearly identical.
\par
Following galaxy-only fits presented in \autoref{sec:ez} and keeping the above considerations in mind, we now would like to explore an AGN-only scenario where we model the observed light with a two component AGN model. The first one is the empirical model based on a composite of 2200 SDSS quasar spectra \citep{vandenberk01}, and the second is derived from 27 near-infrared quasar spectra by \citet{glikman06}. We then combine and renormalize both templates, allowing us to cover the full range from rest-UV to the near-infrared. 
\par
The same approach was already successfully employed in \citet{labbe23} for a photometrically selected sample of red dots, and then later for PRISM spectra of 14 such objects in \citet{greene23} and \citet{kokorev23c}. We fit the unreddened AGN component together with the Small Magellanic Cloud (SMC) law \citep{gordon03} attenuated ($A_{\rm V}$= 0.1 -- 4) version of the same composite template. With the photometric redshift being fixed, we are fitting for a total of three free parameters.
\par
We find the AGN-only fits to be a marginally better representation of the observed photometry, when compared to galaxy-only \textssc{EAZY} fits, with $\langle \chi^{2}_{\nu} \rangle=3.0^{+3.7}_{-1.8}$ for the former and $\langle \chi^{2}_{\nu} \rangle=4.2^{+6.6}_{-1.9}$ for the latter, with a difference of approximately  $\langle \Delta \chi^{2}_{\nu} \rangle \sim 1$. Similar findings were also presented in \citet{labbe23}, even without ALMA photometry, and \citet{barro23}, where no significant $\chi^{2}$ difference exists between dusty star-formation and reddened AGN models. 

\subsection{Extreme Equivalent Width of Emission Lines}
Before focusing on the final sample of reddened AGN candidates we would like to conduct one final test which concerns the potential presence of strong emission lines, particularly H$\alpha$ in the spectra of LRDs. Empirical quasar templates presented in \citet{vandenberk01}, which we used to fit our objects, generally contain bright AGN with a rest-frame H$\alpha$ EW$\sim190$ \AA. Conversely, the recent literature results which analyze LRD spectra \citep{killi23,matthee23} have found that the EW of H$\alpha$ can reach and even exceed 500 \AA. Such strong emission lines can contribute to the flux observed in the medium and even broad-band JWST filters in a non-negligible way, making the observed colors redder. In return, if such strong emission lines are not present in the templates, the value of the $A_{\rm V}$, and subsequently other physical properties dependent on it (e.g. $L_{\rm bol}$) can be overestimated.

To test the significance of this effect we do the following. Starting with the original combined \citet{vandenberk01} and \citet{glikman06} template set, we isolate the regions that cover the H$\beta$+[OIII] and H$\alpha$ lines, and use a spline function to fit the continuum, while masking out the regions containing line complexes. While doing this we successfully verify that the measured rest-frame EW of these lines is exactly as the one reported in \citet{vandenberk01}. Finally, we uniformly boost the continuum subtracted spectrum to a point where the EW of the H$\alpha$ line measures at $\sim500$ \AA, and add back the continuum. We then re-fit all of our sources, following the same considerations as described in \autoref{subsec:quasar_fit}.

Using models with boosted emission line strengths, we indeed find the best-fit $A_{\rm V}$ values to be systematically lower, albeit only by $\sim0.1$ mag, on average, compared to the original templates. This offset is well within our quoted uncertainty on the $A_{\rm V}$ from the SED fitting. We thus conclude that even if some of our AGN candidates indeed contained very high EW H$\alpha$ emission lines, the physical properties derived with the original \citet{vandenberk01} template set should still remain valid. Despite being small, this offset is systematic, so we still incorporate it into out uncertainties when computing number densities in the subsequent sections.

\subsection{Final Sample of Little Red Dots}
Following the initial object selection and SED fitting we are now in a position to define our final sample of ``little red dots''. 
The primary goal of this work is to explore the photometrically selected dusty AGN candidates in the high-$z$ Universe, compare these results to robust samples of spectroscopically identified BL AGN, and potentially extend these examinations to fainter UV magnitudes and bolometric luminosities. The accurate determination of these parameters is contingent upon good coverage of the spectral break between the blue and red components at $\sim 4000$ \AA. This is crucial to confirm that the selected objects indeed exhibit the characteristic features of LRDs. Furthermore, a thorough sampling of the rest-frame UV around $\sim1450$ \AA\, is essential to accurately derive $M_{\rm UV}$, and the $5100$ \AA\, rest-frame optical continuum is needed for determining the bolometric luminosity - $L_{\rm bol}$. With the exception of CEERS, all of our fields benefit from full NIRCam filter coverage, spanning from F090W to F444W, which will cover the rest-frame UV at $z\gtrsim4$.  On the other hand CEERS has extremely deep ($\sim 29.6$ mag at $5\sigma$) \textit{HST}/ACS F814W coverage instead, which will also allow us to adequately compute $M_{\rm UV}$ at 1450 \AA\, in the same redshift range. We thus limit our exploration only to objects which have $z>4$. To do that we take into account the $p(z)$ and ensure that the 16th percentile, rather than just the median of the $p(z)$ lies above our redshift threshold \citep[e.g. see][]{valentino23}. This final selection leaves us with a total of 260 red dots.

\subsection{Physical Parameters}

The physical sizes of objects in our final sample are extremely compact, with a median effective radius of $r_{\rm eff}<130$ pc (95 \% upper limit). This is much smaller when compared to the typical rest-optical sizes of star-forming galaxies measured at $z>5$ \citep[e.g. see][]{kartaltepe23,ormerod24}, but is similar to the extremely compact red objects presented in \citet{labbe23_nat,labbe23,baggen23} and LRDs spectroscopically confirmed as BL AGN  \citep{furtak23_nat,kokorev23c}. Curiously, dusty galaxies at $z>7$ explored in \citet{akins23} also show a lack of extended bright component ($r_{\rm eff}<200$ pc), similar to LRDs. Although not as faint or centrally concentrated as our objects or other LRDs at these redshifts, some of these similarities might imply that these dusty objects can act as potential AGN hosts.
\par
Using the standard relations, with the scatter, presented in \citet{greene05} and taking into account our best-fit $A_{\rm V}$ ($\sim$ 0.6 -- 3.7 mags), we derive the $L_{\rm bol}$ from the 5100 \AA\, continuum, measured directly from best-fit SEDs. While this is not ideal, and assumes that the red continuum is AGN dominated, the SED model-dependent values represent our best guess for the intrinsic AGN luminosities. The inferred bolometric luminosities for the compact red objects from our sample thus range from $L_{\rm bol}\simeq 10^{43.5}-10^{46.5}$ erg/s. This range is slightly brighter than that derived in \citet{labbe23} as we are not including any lensed fields, and thus likely fail to detect intrinsically fainter LRDs. We show the dust-corrected and observed $L_{\rm bol}$ values in \autoref{fig:fig2}.
\par
In \autoref{fig:fig3} we explore how the observed $M_{\rm UV}$ values of our LRDs compare to the expectations derived from the dust corrected bolometric luminosity \citep{shen20}. Given our median $\langle A_{\rm V}\rangle\sim1.6$, we expect the UV extinction to be large with $A_{\rm UV}\sim9$, however what we find is $\langle A_{\rm UV} \rangle \sim 2.5$ \citep[similar to e.g.][]{greene23,maiolino23b,matthee23}, more than six magnitudes difference. Adding to this, the shape of the rest-UV spectrum, while faint, does not hint at any dust extinction. This suggests that a second component, different from a reddened AGN spectrum is present in LRDs, however with our current data its origin can not be determined.
\par
The final table which contains photometry, sizes and the physical parameters we derive for our sample is available in full online \footnote{\url{https://github.com/VasilyKokorev/lrd_phot}}. We show an excerpt of the full table in the Appendix.

\begin{figure*}
\begin{center}
\includegraphics[width=.95\textwidth]{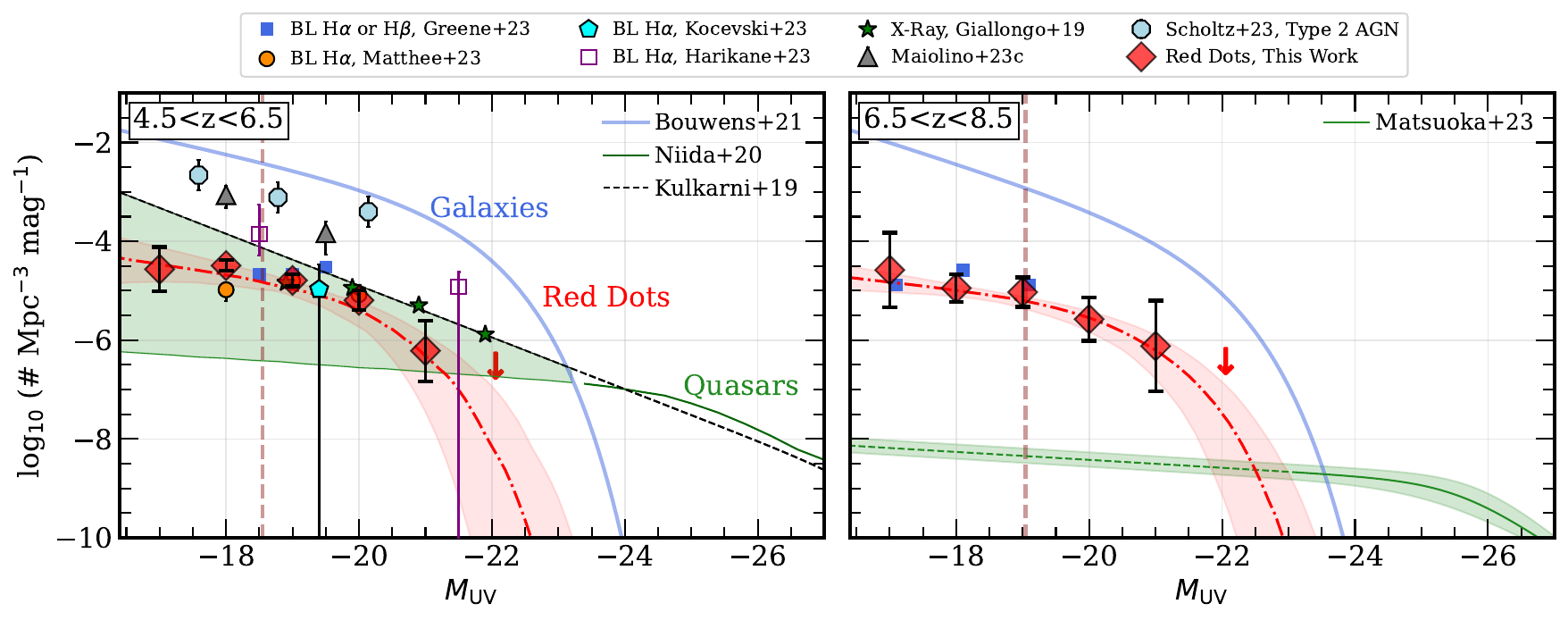}
\caption{The UV luminosity function (UVLF) for the LRDs in our sample in the z$\in[4.5,6.5]$ (left) and z$\in[6.5,8.5]$ (right) bins (blue), derived from observed rest-UV light. Upper limits are shown is downward pointing arrows. The dashed red line and the shaded area correspond to our best-fit Schechter function and its 68 \% confidence interval, respectively. Vertical maroon lines highlight the $M_{\rm UV}$ completeness limit calculated from the average depth of F814W/F090W bands. We compare our derived number densities to the luminosity functions of Lyman break galaxies from \citet{bouwens21} (solid blue line), extrapolated quasar UVLF relations from \citet{niida20} at $z\sim5$, as well as an upper bound provided by \citet{kulkarni19} (green lines). At $z\sim7$, in green, we show the UVLF derived from bright quasars from \citet{matsuoka23}. We highlight the spectroscopically identified LRDs from \citet{greene23} (blue squares), \citet{matthee23} (orange circles) and \citet{kocevski23} (blue pentagon). Furthermore, we show densities of BL AGN quasars from \citet{maiolino23b} (gray triangles) and \citet{harikane23_agn} (open squares). Green stars show the UV number densities of the X-ray detected quasars at $z\sim5$ from \citet{giallongo19}. Finally, light blue octagons represent the UVLF derived for galaxies hosting Type 2 AGN from \citet{scholtz23}. We offset some of the literature points by $\pm0.05$ dex horizontally for visualization purposes. Note that our measured UV luminosities do not decompose the AGN emission from the potential galaxy light.}
\label{fig:fig4}
\end{center}
\end{figure*}

\section{The Number Density of Compact Red Sources}
\subsection{Estimating Effective Volumes} \label{sec:eff_vol}

One of the key motivations for our work is to conduct an unbiased search for LRDs in some of the deepest blank fields observed with \textit{JWST}. Our goal is to extend the existing luminosity functions that have been deduced from spectroscopic samples, for a larger sample covering a wider area. By applying the color and size criteria that have been adopted in recent LRD studies, such as those discussed in \citet{greene23} and \citet{labbe23}, we aim to exploit the large area in our blank fields to get better statistics, particularly at the bright and faint ends.
This approach should allow us to examine how much these objects contribute to the observed $L_{\rm bol}$ and $M_{\rm UV}$ number densities. However, as we are working with a photometrically selected sample our analysis will be focused on the aggregate characteristics of the LRDs, rather than on detailed examinations of individual objects.

Focusing only on the blank fields allows us to estimate the effective volumes for our objects in a rather simple way. In order to measure the observed number densities of our sample, we follow the standard $V_{\rm max}$ method \citep{schmidt}. The $1/V_{\rm max}$ estimator has the advantage of simplicity and does not require prior assumptions on the functional form for the luminosity distribution, ideal for LRDs since their intrinsic luminosity/mass distributions are unknown. To compute the number density for some property - $x$, we can then say:
\begin{equation*}
\Phi(x) = \frac{1}{\Delta x} \sum_i V_{\rm max,i}(A,z_{\rm min},z_{\rm max})^{-1},
\end{equation*}
where $\Delta x$ is the width of the bin and $V_{\rm max,i}$ is the maximum volume over which a source can be detected. In return, $V_{\rm max,i}$ depends on the effective survey area - $A$, lower redshift bin boundary - $z_{\rm min}$ and maximum observable redshift - $z_{\rm max}$. The latter is computed empirically from the detection limits of the survey, given the selection criteria, and cannot exceed the maximum redshift of the bin. 

We obtain the total survey areas by adding up all the non-masked pixels in our detection images, as presented in \autoref{tab:tab1}. Given how bright we require our objects to be (F444W$<27.7$ mags at SN$>14$) it might seem that $z_{\rm max}$ would always exceed the maximum redshift of the bin, however this does not take into account the fact that our objects have to be detected in at least four bands (at $>3\sigma$) to make color selection robust. We choose to remain conservative with our volume corrections, by only requiring one band per color combination to be detected. The $z_{\rm max}$ values for each object are then estimated by considering our color selection laid out in \autoref{sec:ident}. Uncertainties on our number densities are then derived in the following way. We consider the standard errors arising from Poisson statistics and compute them as prescribed in \citet{gehrels86}. Given that we only consider a photometric sample in our work, the uncertainty on the photometric redshift has to be taken into account appropriately in order to derive realistic errors on the physical parameters and number densities. To do that we follow the approach described in \citet{marchesini09}. Briefly, for each object we use Monte Carlo simulations to determine whether the objects fall into the redshift bin by considering their $p(z)$. The final uncertainties are then a quadrature sum of the Poisson and $p(z)$ errors.

Accounting for magnitude incompleteness effects as it is normally done for galaxy luminosity functions is not possible in our case, since it relies on making assumptions regarding the intrinsic source distributions. However, as \citet{labbe23} already note, the requirement for objects to be bright in the detection band should lessen, but not eliminate altogether, the effect of magnitude incompleteness. Given that our sources are compact, we also expect that all of them will be detected above the brightness limit, diminishing the need to consider the incompleteness as a function of surface brightness. Despite the complex selection function, it is still possible to define a limit beyond which the derived number densities, for observed quantities, are expected to become incomplete. We will discuss this in the next sections.

\subsection{UV Luminosity Function}
In \autoref{fig:fig4} we present the UV luminosity functions in two redshift bins, at $z\sim5$ and  $z\sim7$, derived from the continuum luminosity at rest frame $1450$ \AA\, as normally done for blue quasars. We list the number counts alongside the uncertainties in \autoref{tab:tab2}. The widths of our redshift bins were chosen to best align with the current literature results, for ease of comparison, as well as to ensure that photometric redshift uncertainties have a minimal impact on the luminosity functions.
\par
At $z\sim5$, we find that the number densities of our red color selected AGN are $\sim2$ dex higher compared to the UV-selected quasars at similar magnitudes, depending on the extrapolation \citep{niida20}. As an upper limit on number density of quasars at $z\sim5$, we also compare to the results presented in \citet{kulkarni19}, which combine both UV-bright quasars ($M_{\rm UV}<-24$) and UV-faint X-ray detected AGN \citep{giallongo19} in their UVLF. 
\par
Before comparing to current observational results \citep{kocevski23,greene23,matthee23}, we note that it is difficult to accurately define the selection function for spectroscopically observed samples, and therefore derive the $V_{\rm max}$ corrections. As such the number densities computed in these works should be treated as lower-limits. In our case the sample is selected via photometry and we derive our $V_{\rm max}$ correction based on the selection criteria alone. This is done to avoid significantly over-estimating the number counts, thus misrepresenting the true abundance of red dots. It is also unlikely that this difference is a result of brown dwarfs contaminating our sample here, since we introduce an additional color cut from \citet{greene23} based on the spectra from \citet{burgasser23}.
\par
Taking the uncertainties into account, we find that our UV number counts are consistent with \textit{JWST}-selected red BL AGN samples \citep{greene23,labbe23,matthee23}, at least at $M_{\rm UV}\sim-19$ and brighter. Confirming the initial findings for the UNCOVER red-dots presented in \citet{greene23} and \citet{labbe23}, we also find that our sample accounts for $\sim 10$ -- 30 \% of total BL AGN populations at high-$z$ \citep{harikane23_agn,maiolino23b} and is largely comparable to the X-ray selected quasars from \citet{giallongo19}, although in the case of the latter we infer higher number densities at fainter UV magnitudes. However it is worth noting that differences between the resolution of Chandra X-ray data and optical light from \textit{HST} can lead to uncertainties when associating X-ray emission to the galaxies being present in the same patch of the sky. Curiously enough, the recovered scarcity of compact red sources compared to galaxies is in stark contrast to the density of Type 2 AGN hosts inferred from the recent JADES spectra \citep{scholtz23} which report as much as a $20 \%$ contribution to the galaxy luminosity functions at $z\sim5$.
\par
When moving to the $z\sim7$ bin, the results for the UVLF at both bright and faint luminosities are inconclusive, due to the limited number of objects and the uncertainty on the photometric redshifts. However, we are again consistent with the number densities of UNCOVER BL AGN from \citeauthor{greene23}. Comparing to the luminosity functions of UV selected quasars from \citet{matsuoka23} at $z\sim7$, and extrapolating to fainter magnitudes, we find a $2-3$ dex offset between the number densities at $M_{\rm UV}>-22$, roughly a factor of ten larger than in the lower redshift bin.
Alongside our UVLF we also highlight the median $M_{\rm UV}$ $5\sigma$ completeness limits. This is derived by considering the depths of filters covering the rest frame $\sim1450$ \AA\, at a given redshift, and whether a source of a given $M_{\rm UV}$ would be detected at a S/N$>5$. As such we should be complete down to $M_{\rm UV}\sim-18.5$ at $z=5$, and $M_{\rm UV}\sim-19.0$ at $z=7$.
\par
Following \citet{bouwens15}, we fit our observed UV number densities with a Schechter \citep{schechter} function, allowing all parameters to be free. We only fit data brighter than $M_{\rm UV}=-18.5$ at $z\simeq5$ and $M_{\rm UV}=-19.0$ at $z\simeq7$ as our number densities indicate that we are likely becoming incomplete at such faint magnitudes. The best-fit is shown in \autoref{fig:fig4} and the parameters are listed in \autoref{tab:tab_lf_params}. In both redshift bins we find that our red compact objects constitute roughly $3-5 \%$ of the total star-forming galaxy populations \citep{bouwens21}, consistent with the spectroscopic samples of red-dots \citep{greene23}.
We also report shallower faint-end slopes compared to SF galaxies, however it is likely that the observed flattening of the UVLF for LRDs is induced by the incompleteness of our sample at fainter UV magnitudes, rather than any lack of compact red sources at fainter UV magnitudes. Deeper surveys would be required to robustly constrain the faint-end slope of LRDs. It appears that the LRD luminosities start to become comparable or even outshine galaxies at brighter ($M_{\rm UV}\sim -23$) magnitudes, which is particularly prominent in the $z\sim7$ bin. This might be an expected consequence of the assembly of increasingly massive black holes with cosmic time \citep[e.g. see][]{piana22}, or selection effects \citep[see][]{volonteri17}, however we note that our number counts for the brightest objects are uncertain due to a limited amount of detections available.
\par
Provided that our color and morphology selection is 
comparably successful at identifying reddened AGN as was previously shown \citep{labbe23,greene23}, it appears that the compact red sources identified in blank \textit{JWST} fields are $\sim 1 - 2$ dex more numerous compared to the pre-\textit{JWST} studies of known UV-selected faint quasars ($M_{\rm UV}>-21$). While this trend has been consistently re-emerging in the new \textit{JWST} results \citep[e.g.][]{furtak23_nat,kokorev23c,maiolino23b,pacucci23}, it is worth noting that earlier works have already hinted that the number density of UV faint, dusty active black holes could have been much higher than previously thought \citep{laporte17,morishita20,fujimoto22}. For example, both \citet{fujimoto22} and \citet{morishita20} find that the less-luminous red quasar population could be anywhere from 10 to 100 times more common at $z\sim7-8$, compared to quasars luminosity functions at $z\sim6$, constructed from ground-based datasets \citep[e.g.][]{matsuoka18,kato20,niida20}. The results of this work, together with the recent efforts to study compact red sources, therefore imply that these faint quasar populations, missed by previous surveys, are now being uncovered by the deep and rich multi-wavelength photometry and spectra from \textit{JWST}. It is also important to highlight that if we extrapolate our UVLF to brighter magnitudes, the number density of LRDs becomes comparable to and then drops below the density of UV-selected quasars. Currently, however, it is not possible to speculate whether this is a real physical effect, or simply a consequence of insufficient volumes sampled.

\begin{figure*}
\begin{center}
\includegraphics[width=.78\textwidth]{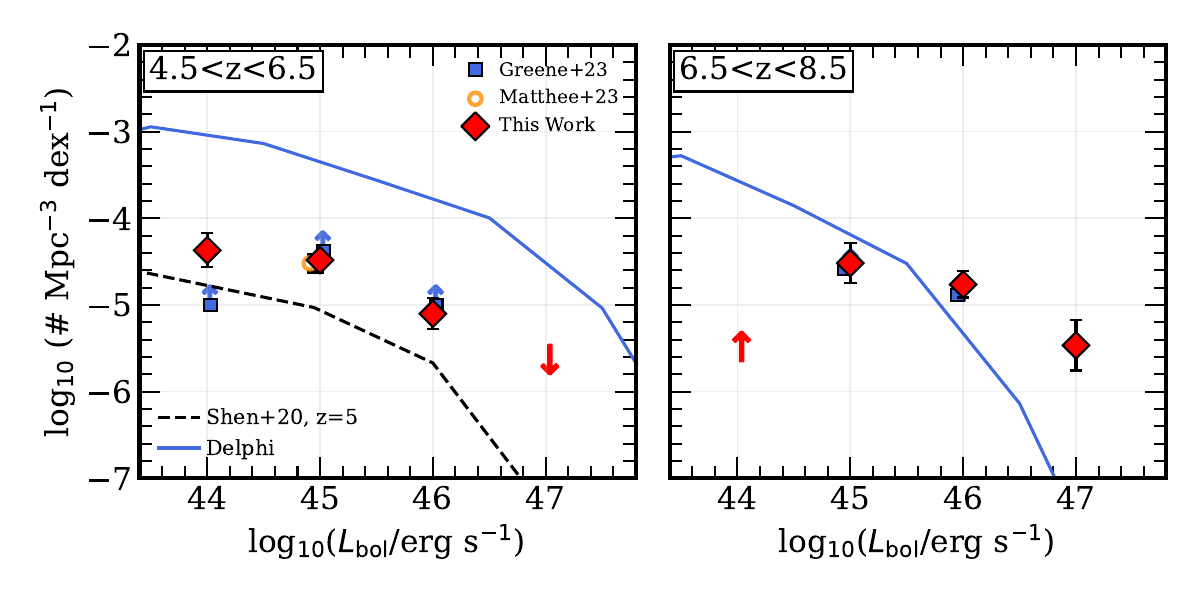}
\caption{Bolometric luminosity functions in the z$\in[4.5,6.5]$ (left) and z$\in[6.5,8.5]$ (right) bins, derived from $L_{\rm 5100}$, assuming rest-frame optical continuum is AGN dominated. The number densities have been $V_{\rm max}$ and completeness corrected. Uncertainties are derived from Poisson noise \citep{gehrels86}. Arrows show upper limits on the derived number densities. Blue squares show upper limits derived for spectroscopically confirmed ``little red dots'' in the UNCOVER data from \citet{greene23}. In addition in the lowest redshift bin we show the NIRCam grism result of \citet{matthee23} (open circle).
Dashed lines show the pre-\textit{JWST} $L_{\rm bol}$ relation derived in \citet{shen20}. Finally, the blue lines show the luminosity function from the Delphi semi-analytic models \citep{dayal19} that grow SMBHs from seeds.}
\label{fig:fig5}
\end{center}
\end{figure*}

\subsection{Bolometric Luminosity Function}

Our SED fitting results show that the fraction of the UV light contributing to the total luminosity is small as a result of significant dust reddening ($A_{\rm V}=0.6$--2.7) in these objects. Even with spectra in hand \citep[e.g.][]{greene23}, it is not easy to establish the origins of the rest UV light, which could be AGN light, either scattered or transmitted through patchy dust clouds, or unobscured light from star-formation in the host galaxy. As such, while we put our LRDs in the context of their observed UV luminosities, this does not explicitly describe the physics of potential AGN these compact objects host. Due to that, and also to carry out a comparison with existing spectroscopic bolometric luminosity functions of dusty BL AGN, we also present bolometric luminosity functions in \autoref{fig:fig5} and \autoref{tab:tab2}. While dust attenuation, estimated from SED fitting, can be an important source of uncertainty, we note that even if all our $A_{\rm V}$ values were grossly overestimated, this would only change the dust-corrected $L_{\rm bol}$ by a factor of $\times5$, given the average $A_{\rm V}\sim1.6$. This would thus only impact the number densities by $\sim\sqrt{5}$ on average, which is insignificant when compared to the Poisson and $z_{\rm phot}$ errors. Finally, to account for the potential presence of emission lines with high EW, we incorporate the additional $\sim0.1$ systematic shift in $A_{\rm V}$ and apply it to the uncertainties on the $L_{\rm bol}$.
\par
Understanding where the bolometric luminosity functions start to become incomplete is less straightforward compared to the observed quantities like $M_{\rm UV}$, as the former also relies on dust correction derived via SED modeling, and the assumptions made regarding the AGN contribution to the rest-frame optical emission. For this reason we do not define a completeness cut, like we do for $M_{\rm UV}$, however for each bin of bolometric luminosity with a width of 1 dex, we also compute the $V_{\rm max}$ correction as was described in \autoref{sec:eff_vol}.

Our number densities again confirm that the red-compact AGN candidates are roughly 100 times more abundant compared to the UV-selected AGN at similar intrinsic luminosities \citep{shen20} at $z\sim5$. The $L_{\rm bol}$ number densities which we recover are comparable to the previous results for these objects derived in \citet{greene23}, \citet{labbe23} and \citet{matthee23} for $L_{\rm bol}-10^{45-46}$ erg/s. Curiously however, we find a factor of ten more LRDs compared to \citet{greene23} at $L_{\rm bol}\sim10^{44}$~erg/s. Nominally, the median NIRSpec depth at 4 $\mu$m of the UNCOVER follow-up of Abell 2744 is shallower compared to the fields we examine, as such it is perhaps unsurprising that we can recover a large fraction of intrinsically faint objects. However since the $L_{\rm bol}$ is not an observed quantity and depends on SED modeling to calculate the dust correction it is difficult to ascertain whether the higher number densities we recover are indeed caused by the depth difference, or simply the bias caused by the spectroscopic-only sample selection and lensed volumes in UNCOVER. Moreover, the mask design of the UNCOVER NIRSpec observations in the Abell 2744 field was also driven by optimizing the MSA coverage to include other targets of interest and was not just limited to LRDs. This, in return, induces selection effects which would not be possible to trace back and correct for.
\par
We additionally compare our bolometric luminosity function to the latest version of the semi-analytic \textsc{Delphi} models \citep{dayal19,dayal24}. In brief, these models follow the seeding and growth of BHs from $z\sim40$ down to $z\sim4.5$. Included are also all the key processes of merger- and accretion driven assembly of dark matter halos and their baryonic component (including black holes). The model also follows star formation and black hole growth and their respective feedbacks in determining the assembly of these early systems. Finally, \textsc{Delphi} models also include key dust processes to yield dust-to-stellar mass ratios, which with a baseline constructed against the latest ALMA observations \citep{dayal22,mauerhofer23}. All of this was specifically done to ensure \textsc{Delphi} can reproduce both the intrinsically faint and reddened sources in the recent literature, i.e. the LRDs. 

We find that while our observations are comparable to \textsc{Delphi} results at $L_{\rm bol}<10^{47}$ erg/s at $z\sim7$, these models fail to reproduce the high number density of bright objects we report. At $z\sim5$ on the other hand, our densities consistently fall 1 dex below \textsc{Delphi} predictions. This in return could suggest that the fraction of dusty AGN is diminishing toward later times, as they potentially transition to unobscured quasars \citep[][]{fu17,fujimoto22}.
\par
Finally, we also see a higher prevalence of intrinsically brighter objects at $\sim L_{\rm bol}-10^{47}$, which is likely consequence of larger volumes sampled in our analysis. As already mentioned in \citet{greene23} however, it is worth noting that the uncertainties on the $L_{\rm bol}-L_{5100}$ relation, dust correction and assuming that these objects are dominated by AGN light at rest-frame optical could cause objects to scatter upwards into the high luminosity bins. We only recover a single object above $L_{\rm bol}=10^{47}$ erg/s at $z\sim5$ and below $L_{\rm bol}=10^{44}$ erg/s at $z\sim7$, respectively. As this is insufficient to properly compute luminosity functions, these are shown as upper (lower) limits in \autoref{fig:fig5} derived by combining the Poisson \citep{gehrels86} and photo-$z$ \citep{marchesini09} uncertainties.

\begin{figure}
\begin{center}
\includegraphics[width=.49\textwidth]{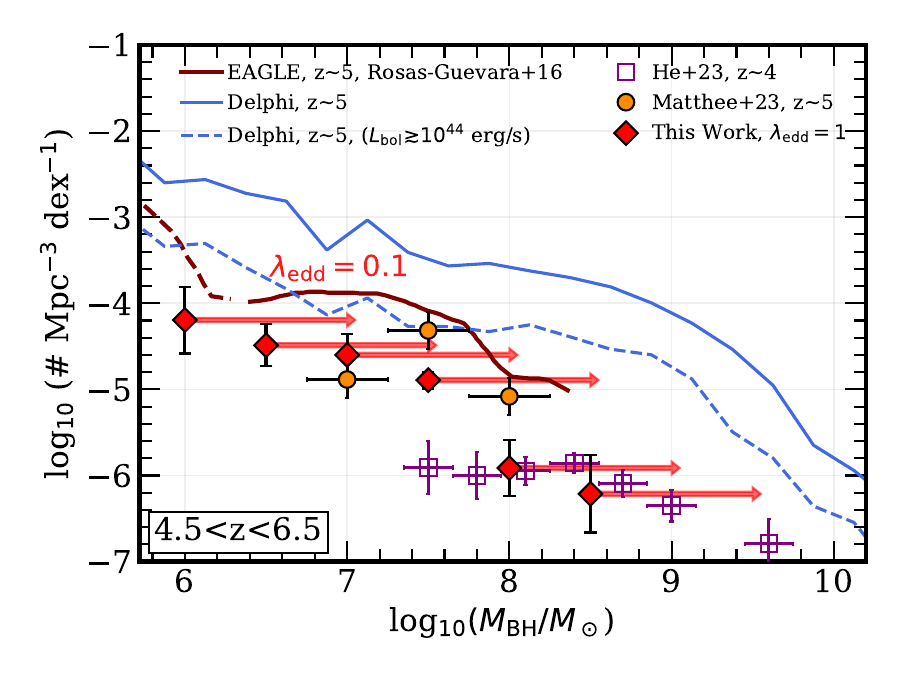}
\caption{The SMBH mass function, assuming $\lambda_{\rm edd}=1$, of our sample in the $4.5<z<6.5$ range. Red arrows show how our mass function would change, if we assumed a lower Eddington ratio of 10 \% . We overlay the SMBH mass function from \citet{matthee23} at $z\sim5$ in orange and HSC+SDSS derived BH mass function from \citep{he23} in magenta. The maroon line shows the results from the EAGLE simulation at $z\sim5$ \citep{rosas-guevara16}. The solid and dashed blue lines show the result from Delphi \citep{dayal14,dayal19,dayal20} simulations for all and bright ($L_{\rm bol}\gtrsim10^{44}$ erg/s) black holes, respectively. Measured number densities of our LRDs agree well with the spectroscopic sample from \citep{matthee23} and simulations given a $\lambda_{\rm edd}\sim1$. }
\label{fig:fig6}
\end{center}
\end{figure}

\begin{deluxetable}{ccc}
\tabcolsep=0.3cm
    \tablecaption{Bolometric and UV ($\lambda_{\rm rest}=$1450 \AA\,) luminosity functions, as well as a SMBH mass function for our sample of LRDs.\label{tab:tab2}}
    \tablehead{\multicolumn{3}{c}{UV Luminosity}}
    \startdata
$M_{\rm UV}$ [ABmag] & N & $\Phi$ / [cMpc$^{-3}$ mag$^{-1}$] \\
\hline
\multicolumn{3}{c}{$4.5<z<6.5$ \hfill} \\
\hline
-17.0 & 19 & $-4.56\pm0.45$ \\ 
-18.0 & 68 & $-4.49\pm0.11$ \\ 
-19.0 & 53 & $-4.79\pm0.12$  \\ 
-20.0 & 21 & $-5.19\pm0.20$  \\ 
-21.0 & 2 & $-6.21\pm0.61$  \\ 
-22.0 & 1 & $<-6.52$  \\ 
\hline 
\multicolumn{3}{c}{$6.5<z<8.5$ \hfill}\\
\hline
-17.0 & 5 & $-4.58\pm0.75$ \\ 
-18.0 & 23 & $-4.95\pm0.28$ \\ 
-19.0 & 29 & $-5.02\pm0.30$ \\ 
-20.0 & 9 & $-5.58\pm0.44$ \\ 
-21.0 & 2 & $-6.12\pm0.92$ \\ 
-22.0 & 1 & $<-6.42$  \\ 
\hline
\hline  
\multicolumn{3}{c}{Bolometric Luminosity}\\
\hline
log$_{10}$($L_{\rm bol}$/erg s$^{-1}$) & N & $\Phi$ / [cMpc$^{-3}$ dex$^{-1}$] \\
\hline 
\multicolumn{3}{c}{$4.5<z<6.5$ \hfill} \\
\hline
44.0 & 46 & $-4.37\pm0.20$ \\ 
45.0 & 93 & $-4.48\pm0.10$ \\ 
46.0 & 26 & $-5.10\pm0.18$ \\ 
47.0 & 1 & $<-5.62$ \\ 
\hline 
\multicolumn{3}{c}{$6.5<z<8.5$ \hfill}\\
\hline
44.0 & 1 & $>-5.48$ \\ 
45.0 & 14 & $-4.51\pm0.23$ \\ 
46.0 & 33 & $-4.76\pm0.15$ \\ 
47.0 & 9 & $-5.47\pm0.29$ \\ 
\hline
\hline
\multicolumn{3}{c}{Black Hole Mass ($\lambda_{\rm edd}=1$)}\\
\hline
log$_{10}$($M_{\rm BH}$/$M_\odot$) & N & $\Phi$ / [cMpc$^{-3}$ dex$^{-1}$] \\
\hline
\multicolumn{3}{c}{$4.5<z<6.5$ \hfill} \\
\hline
6.0 & 39 & $-4.19\pm0.39$ \\
6.5 & 44 & $-4.40\pm0.24$  \\
7.0 & 40 & $-4.60\pm0.24$  \\
7.5 & 21 & $-4.89\pm0.10$  \\
8.0 & 2 & $-5.91\pm0.33$  \\
8.5 & 1 & $-6.21\pm0.45$  \\
\enddata
\end{deluxetable}

\subsection{The $z\sim5$ SMBH Mass Function}
With the data we have obtained, we will now derive and describe the measurement for the supermassive BH mass function which our compact objects potentially host. Computing the mass of the central black hole generally requires knowledge of the width of the broad lines (e.g. H$\alpha$, H$\beta$ in rest-optical or Mg II in rest-UV) coupled with the luminosity of their broad components or the luminosity derived from the AGN continuum at $\lambda_{\rm rest}=5100$ \AA\, \citep[see e.g.][]{kaspi00,greene05}. 
\par
While secure determination of the black hole mass in our compact objects is not possible due the photometric nature of the sample, we can still place a lower limit on black hole masses ($M_{\rm BH}$), by making a set of conservative assumptions. To do that, we adopt a scenario where all our AGN candidates accrete at Eddington rate (the physical limit at which outward radiation pressure balances inward gravitational force), such that $L_{\rm bol} \sim L_{\rm edd}$, where $L_{\rm edd}$ is directly proportional to $M_{\rm BH}$.  While in the literature the Eddington rate ($\lambda_{\rm edd}$) for confirmed AGN in LRDs was found to vary between $10-40$ \% \citep{furtak23_nat,greene23,kokorev23c}, we would like to remain conservative and compute a lower limit on the $M_{\rm BH}$. It is also worth noting that, given this range of $\lambda_{\rm edd}$ in the literature, this is still small compared to other sources of uncertainty in our work.
\par
We calculate the $M_{\rm BH}$ directly from the dust-corrected $L_{\rm bol}$ and compute the SMBH mass function as described in the previous sections. We present our SMBH mass function in \autoref{fig:fig6} and \autoref{tab:tab2}, binned into 0.5 dex intervals to allow a direct comparison with the existing observational and theoretical results in this redshift range. We limit this investigation to the $z\sim5$ range only. As before, we note that the effect of the $A_{\rm V}$ uncertainty on our number densities is expected to be at most $\sim \sqrt{5}$, largely overshadowed by the Poisson and redshift errors.
\par
We are now in a position to compare our mass function to the existing samples of both bright and faint quasars at $z\sim5$. We start with the latest ground based examination of the quasar mass function at $z\sim4$ from \citet{he23}. The authors focus on a sample of $\sim1500$ faint broad-line AGN, from a combined Hyper Suprime Cam (HSC) and SDSS dataset, allowing them to extend their examination to a low mass range we are most interested in ($M_{\rm BH}\simeq10^{7-8}$ $M_\odot$). We find that, while our result is consistent with the ground based mass function in the high mass regime $M_{\rm BH}>10^{8}$ $M_\odot$, our number densities diverge below that mass and continue to rise up to $\sim10^{-4}$ cMpc$^{-3}$ at $M_{\rm BH}\simeq10^{6}$ $M_\odot$. Barring the color selection, it is possible that this effect is purely observational, as the SDSS/HSC detection limits in the rest-UV are much shallower compared to the \textit{JWST} fields we explore.
\par
Furthermore, we contrast our result to the BH mass function at $z\sim5$ based on a sample of LRDs from the slitless \textit{JWST} derived in \citet{matthee23}. Given the uncertainties, we find our results to be consistent within $1\sigma$, although we do not find a sharp drop-off in number densities at $M_{\rm BH}<10^7$ $M_\odot$, likely driven by low mass incompleteness of grism data as mentioned in \citeauthor{matthee23}. 
\par
Naturally, the fact that both our result and \citet{matthee23} find more low mass black holes compared to \citet{he23} is unsurprising, given the depth and wavelength coverage of \textit{JWST} data.
Quite curiously however, it appears that the mass function derived from dusty compact LRDs seems to
nicely continue the rising trend of ground-based data, and extend the SMBH mass functions toward $M_{\rm BH}\sim10^6$ $M_\odot$. In this redshift range, the maximum volume sampled by our multi-field investigation is roughly equal to $\sim 3.0\times10^6$ cMpc$^3$. Therefore, taking into account the results of \citet{he23}, we should expect only one object with $M_{\rm BH}\sim10^{8.5}$ $M_\odot$ in our images, which indeed is the case. Detection of AGN hosting black hole with masses larger than that, would, however, require survey sizes ten to twenty times larger.  
\par
Before drawing conclusions, we would like to conduct a final check, and compare our result to the hydro-dynamical simulation EAGLE \citep{rosas-guevara16} and semi-analytic Delphi \citep{dayal14,dayal19,dayal20,dayal24} simulations describing masses of SMBHs in the same redshift range. We limit our examination of the Delphi models to the bright ($L_{\rm bol}\gtrsim10^{44}$ erg/s) regime to match the same luminosity range covered by our objects. In the intermediate, to low mass end ($M_{\rm BH}<10^{7.5}$ $M_\odot$) our results are in broad ($\sim 2\sigma$) agreement with both EAGLE and Delphi, however in both cases we start to see a significant difference in number densities as we move to higher masses. Perhaps a worthwhile question to ask in this case, is whether more of these high-mass BHs would be found in larger areas, we will discuss this in the later section. 
\par
Examining both the UV and bolometric luminosity functions we note that LRDs only represent $\sim 25$ \% of the total type I (broad-line) AGN population as inferred by \citet{harikane23_agn,maiolino23b}, even less so compared to the most recent examination of type II AGN hosts from JADES \citep{scholtz23}, where LRDs are 30-40 times less numerous. Taking this into account, we can conclude that, at least at $z\sim5$, LRDs appear to represent at most 1 \% of the total accreting BH population over the $L_{\rm bol}\sim10^{44-47}$ erg/s range. The fact that LRDs are truly a distinct population of dusty broad-line AGN can therefore explain the observed $\sim 2$ dex disparity between our results and simulations. Finally we would like to reiterate that our investigation of the BH mass function relies on assuming the most conservative case of accretion at exactly the Eddington rate, as we do not want to erroneously overestimate the number of high-mass black holes. Keeping that in mind, in \autoref{fig:fig6} we also show how our mass functions would change if we were to assume an Eddington ratio of 10 \% instead. In this case we find that while our number densities compared to UV samples are still high, we now more closely match the abundance of high mass SMBHs predicted by \textsc{Delphi}. However, until broad emission line observations for all our sources are available, the value of $\lambda_{\rm edd}$ will remain uncertain.

\begin{deluxetable}{cccc}
    \tablecaption{Best fit Schechter parameters for the rest-frame UVLF at $\lambda_{\rm rest}=1450$ \AA, across blank \jwst\, fields.\label{tab:tab_lf_params}}
    \tablehead{$\langle z \rangle $ &
    $M^*_{1450}$ [ABmag] &
    $\phi^*[10^{-3}$ Mpc$^{-3}$] & 
    \quad$\alpha$}
    \startdata
\hline
5 & $-20.64\pm0.67$ & $0.008\pm0.003$ & $-1.76\pm0.67$\\
7 & $-20.67\pm0.76$ & $0.005\pm0.002$ & $-1.46\pm0.37$ \\
\enddata
\end{deluxetable}

\section{Discussion and Summary} \label{sec:disc_sum}
\subsection{Abundance of bright compact sources}
Previously limited to UV$-$selected samples at $z\lesssim6$ \citep{kashikawa15,banados18,matsuoka18,inayoshi20,wang21,fan22} we are now able to use \textit{JWST} to reveal the presence of AGN during \citep[e.g.][]{kocevski23,matthee23,ubler23} and even beyond the epoch of reionization, \citep[e.g.][]{bogdan23,furtak23,goulding23,kokorev23c,larson23,lambrides23, maiolino23} only hundreds of millions of years after the Big Bang. Standing out among these early studies of active black holes, is the population of reddened type I AGN, the so called ``little red dots'' \citep{greene23,labbe23,matthee23}. 
\par
While the study of this unique population has been mostly limited to small spectroscopic samples, most recent efforts focused on the expansive Abell 2744 \textit{JWST} data-set \citep{labbe23} have shown great promise at using a combination of NIRCam colors and morphology to identify reddened AGN. This initial photometric selection was shown to be remarkably successful with $\sim 80$ \% of targets indeed confirmed as $z>5$ dusty broad-line AGN \citep{fujimoto23_uncover, furtak23, greene23,kokorev23c}. 
It is clear that these objects play an important role 
in the story of black hole growth at early times, however so far a systematic review of these enigmatic AGN across multiple fields has not been undertaken.
\par
Motivated by the success of this photometric selection, we present a sample of 260 reddened BL AGN candidates in the $4<z<9$ redshift range, covering 4 separate blank \jwst\, fields with a total area of $\sim640$ arcmin$^{2}$. We uniformly reduce the NIRCam \jwst\, data from a variety of public programs, complementing our photometric coverage with archival \hst\, observations. We perform a color and morphology selection to identify the most promising compact objects which display a dichotomy in their observed SED shapes, namely a blue rest-UV continuum, and a red power law-like rest-optical. 
\par
Using model fitting, we derive photometric redshifts as well as a range of physical parameters including $A_{\rm V}$, $M_{\rm UV}$ and dust corrected $L_{\rm bol}$. We split our objects into two redshifts bins at $z\sim5$ and $z\sim7$ and explore their contribution to the UV and bolometric luminosity functions of star-forming galaxies, as well as UV-selected quasars. Consistent with the previous works \citep{greene23,matthee23,maiolino23b} exploring high-$z$ BL AGN, we find that number densities of these objects at $z>5$ are surprisingly high, in excess of $\times100$ compared to faint UV selected quasars \citep[e.g.][]{niida20,he23}, while also accounting for $\sim20 \%$ of the total BL AGN population \citep{harikane23_agn,maiolino23b}, and $\sim 1-2$ \% of UV-selected star forming galaxies \citep[e.g.][]{bouwens21}. Moreover, while some of these objects were potentially pinned down as potential sources of reionization in their local environment \citep{fujimoto23_uncover}, it appears that their UV luminosities are still insufficient to contribute to reionization in a significant way \citep{dayal24}. 
\par
Assuming accretion at the Eddington rate, we also place a lower limit on the $M_{\rm BH}$ of our objects, finding that some of these can already be very massive ($M_{\rm BH}>10^7$ $M_\odot$) only a few hundred of years after the Big Bang. Using these masses we were also able to construct our prediction for the SMBH mass function, and for the first time, extend it to the low-mass ($<10^{7}$ $M_\odot$) regime. We find that our mass function results are completely consistent with the number densities derived for faint dusty AGN from \citet{matthee23} at intermediate masses, and are comparable to those from UV-selected samples at high mass \citep{he23}. We note however that while their number densities are similar, the sample presented in \citeauthor{he23} consists of unobscured quasars, and not LRDs, which are thought to be dust obscured AGN. We find that both hydro-dynamical and semi-analytic predictions for the number of black holes at this redshift match our observations below $\sim10^{7.5}$ $M_\odot$, however start to disagree by almost 2 dex at higher masses. These massive and bright black holes are likely to be heavily obscured in the rest-frame UV, and thus are not selected as LRDs due to a lack of a clear blue component.

\vspace{0.82cm}
\subsection{Final Remarks}
Using observed NIR colors to pick out active black holes in extragalactic fields is by no means a novel endeavor and has already been successfully done with the IRAC instrument, onboard the Spitzer Space Telescope \citep{lacy04,stern05,donley12}. This method, however, is still in its infancy when it comes to \textit{JWST} \citep[][]{labbe23,andika24}. As already pointed out by \citet{matthee23}, very few \textit{JWST}
programs that detect dusty AGN, were actually designed with AGN in mind, implying that there is still more that we can to address a growing number of questions about this population.
\par
Firstly, the physical mechanisms that govern BH formation and growth in these systems are still poorly understood, however there exists already a growing body of works which try to decipher this enigmatic population \citep{greene23,silk24}. One such puzzle, is the origin of the blue light found in LRDs. The similarity between blue slopes of low-metallicity star-forming galaxies and quasars does not allow us to make a clear assessment of whether the rest-UV light originates from the AGN itself or the compact host galaxy surrounding it from the continuum alone. One way to solve this is to target the Mg II doublet ($\lambda_{\rm rest}\sim2800$ \AA), the CIV, SiIV or HeII ($\lambda_{\rm rest}\sim1840$ \AA) lines \citep[e.g. see][]{maiolino23} and confirm whether these are broadened or not. This however would require longer integration times with NIRSpec as medium or even high resolution gratings would be required to achieve the necessary spectral fidelity. Moreover, while in principle detection of broad UV emission lines could hint at the AGN being responsible for some UV light, this would not necessarily mean that UV continuum also originates from the same source.
\par
Secondly, there are no models as of yet, which can adequately describe the light we see emerging from these objects. So far, we mainly have had to rely on combinations of dust-free and dust attenuated empirical models of local quasars, which might be adequately describing AGN at high-$z$. Moreover, the lingering uncertainty on the $A_{\rm V}$ correction can 
introduce some biases in our estimates of $L_{\rm bol}$ and the $M_{\rm BH}$. One solution to alleviate this is 
to stack spectra of known LRDs, to define sets of reliable models describing these populations and aiding with further photometric selection.
\par
Thirdly, it is crucial to note that a substantial proportion of massive SMBHs (with $M_{\rm BH}>10^8 M_\odot$) at high redshifts (high-$z$) can be heavily obscured, as implied by \textsc{Delphi}. Similar conclusions were already reached from the X-ray luminosity functions at $z>5$, both from simulations \citep{ni20}, and observations \citep{aird15,vito18}. In addition \citet{trebitsch19} have shown that accreting SMBHs in Lyman break galaxies are rarely UV-bright.
With this in mind, selecting these massive AGN as LRDs would thus not be possible, as a combination of very deep rest-UV imaging and large areas are required. Despite that, these objects should still appear bright in near-infrared, which opens up the possibility of effectively identifying them through large or parallel surveys using MIRI.
\par
Finally, as was already shown recently by \citet{williams23b} and \citet{perezgonzalez24}, MIRI can also assist with clarifying true numbers of AGN among LRDs as some of these could be dusty progenitors of compact ellipticals. 
\par
Early results from \textit{JWST} have already provided us with quite unexpected and remarkable results regarding number densities of early AGN, leading to a shift in our understanding of their formation and growth in the early Universe. Our results highlight the potential of using NIRCam alone to select reddened AGN at high-$z$ in an effort to better understand their properties and abundance. While some limitations to this technique exist, as we already discuss in our work, this provides a crucial set of next steps in order to bridge the gap between UV bright quasars and faint SMBHs. However, it is already evident that the importance of faint, reddened AGN at early times can not be overlooked.


\acknowledgements
We thank the anonymous referee for a number of constructive suggestions, which helped to greatly improve the quality of this manuscript. We are grateful to Dale Kocevski and Kohei Inayoshi for their patience with helping us spot and correct minor inconsistencies in the manuscript. The authors would like to thank Sarah Bosman for insightful discussions about UV-bright quasars at high redshift. We also thank Mauro Giavalisco, Hollis Akins and Meghana Killi for useful discussion regarding the nature of dust obscured AGN. VK and KIC acknowledge funding from the Dutch Research Council (NWO) through the award of the Vici Grant VI.C.212.036. J.E.G. acknowledges support from NSF/AAG grant\# 1007094, and also support from NSF/AAG grant \# 1007052. PD \& MT acknowledge support from the NWO grant 016.VIDI.189.162 (``ODIN"). PD also acknowledges support from the European Commission's and University of Groningen CO-FUND Rosalind Franklin program.  This work is based on observations made with the NASA/ESA/CSA James Webb Space Telescope. The data were obtained from the Mikulski Archive for Space Telescopes at the Space Telescope Science Institute, which is operated by the Association of Universities for Research in Astronomy, Inc., under NASA contract NAS 5-03127 for JWST. All the $JWST$ and $HST$ data used in this paper can be found on MAST: \dataset[10.17909/de9v-7893]{http://dx.doi.org/10.17909/de9v-7893}. Some of the data products presented herein were retrieved from the Dawn JWST Archive (DJA). DJA is an initiative of the Cosmic Dawn Center, which is funded by the Danish National Research Foundation under grant No. 140. TBM was supported by a CIERA Postdoctoral Fellowship. This work used computing resources provided by Northwestern University and the Center for Interdisciplinary Exploration and Research in Astrophysics (CIERA). This research was supported in part through the computational resources and staff contributions provided for the Quest high performance computing facility at Northwestern University which is jointly supported by the Office of the Provost, the Office for Research, and Northwestern University Information Technology.

\software{EAZY \citep{brammer08}, FSPS \citep{conroy09}, pysersic \citep{pasha23}, grizli \citep{grizli}, msaexp \citep{msaexp}.} 

\facilities{\jwst, \hst}

\clearpage

\appendix


\begin{deluxetable*}{ccccccccccccc}[h]
\tabcolsep=0.08cm
    \tablecaption{An example of the table containing all properties of our sources. A\dataset[full version]{https://doi.org/10.5281/zenodo.10820724} of this table is available in the electronic format.
    \label{tab:full_sample}}
    \tablehead{ID  & Field & R.A. & Dec  & \texttt{\{filt\}}\_flux  & \texttt{\{filt\}}\_fluxerr  & $z_{\rm phot}$ & $A_{\rm V}$ & log$ (L_{\rm bol}$/[erg/s]) & $M_{\rm UV}$ & $r_{\rm eff,pix}$ & $r_{\rm eff,phys}$\\
    & & [deg] & [deg] & [$\mu$Jy] & [$\mu$Jy] & & [mag] &  & [ABmag] &  [pix] & [pc]}
    \startdata
\hline
1381 & PRIMER-UDS & 34.440988 & -5.209337 & - & - & $ 5.66 ^{+ 0.03 }_{- 1.66 }$& $ 3.1 \pm 0.3 $ & $ 45.0 \pm 0.1 $ &$ -17.52 \pm 0.3 $ & $ <1.04$ & $<186$ \\
1470 & PRIMER-UDS & 34.260549 & -5.209163 & - & - & $ 5.40 ^{+ 0.06 }_{- 0.09 }$& $ 1.1 \pm 0.1 $ & $ 44.4 \pm 0.2 $ &$ -18.45 \pm 0.1 $ & $<1.23$ & $<230$ \\
1544 & PRIMER-UDS & 34.377413 & -5.209023 & - & - & $ 5.46 ^{+ 0.02 }_{- 0.01 }$& $ 1.0 \pm 0.2 $ & $ 44.5 \pm 0.4 $ &$ -18.84 \pm 0.1 $ & $<1.02$ & $<187$ \\
1939 & PRIMER-UDS & 34.403098 & -5.208186 & - & - & $ 4.36 ^{+ 0.17 }_{- 0.30 }$& $ 1.3 \pm 0.1 $ & $ 44.6 \pm 0.4 $ &$ -17.78 \pm 0.1 $ & $<0.71$ & $<144$ \\
3992 & PRIMER-UDS & 34.367304 & -5.204196 & - & - & $ 5.66 ^{+ 0.02 }_{- 0.02 }$& $ 1.6 \pm 0.2 $ & $ 45.2 \pm 0.2 $ &$ -18.53 \pm 0.1 $ & $<0.55$ & $<104$ \\
7991 & PRIMER-UDS & 34.293687 & -5.196477 & - & - & $ 5.66 ^{+ 1.38 }_{- 0.10 }$& $ 2.4 \pm 0.3 $ & $ 45.9 \pm 0.4 $ &$ -18.47 \pm 0.3 $ &$<1.11$&$<196$ \\
8559 & PRIMER-UDS & 34.465426 & -5.195333 & - & - & $ 7.94 ^{+ 0.09 }_{- 0.13 }$& $ 3.9 \pm 0.1 $ & $ 46.7 \pm 0.2 $ &$ -19.50 \pm 0.1 $ &$<0.68$&$<111$ \\
10404 & PRIMER-UDS & 34.462872 & -5.191960 & - & - & $ 5.66 ^{+ 0.03 }_{- 0.02 }$& $ 2.2 \pm 0.2 $ & $ 45.6 \pm 0.3 $ &$ -18.13 \pm 0.1 $ &$<0.50$&$<101$ \\
11219 & PRIMER-UDS & 34.471126 & -5.190444 & - & - & $ 5.43 ^{+ 0.03 }_{- 0.03 }$& $ 2.7 \pm 0.2 $ & $ 45.9 \pm 0.3 $ &$ -19.20 \pm 0.1 $ &$<0.42$&$<84$ \\
14419 & PRIMER-UDS & 34.399178 & -5.184323 & - & - & $ 4.82 ^{+ 0.27 }_{- 0.07 }$& $ 0.9 \pm 0.2 $ & $ 44.5 \pm 0.3 $ &$ -17.84 \pm 0.1 $ &$<0.64$&$<135$ \\
15257 & PRIMER-UDS & 34.431864 & -5.182780 & - & - & $ 6.92 ^{+ 0.06 }_{- 1.46 }$& $ 1.3 \pm 0.2 $ & $ 45.2 \pm 0.1 $ &$ -18.85 \pm 0.2 $ &$<0.79$&$<133$ \\
16999 & PRIMER-UDS & 34.508058 & -5.180084 & - & - & $ 4.21 ^{+ 0.15 }_{- 0.17 }$& $ 1.1 \pm 0.1 $ & $ 45.2 \pm 0.2 $ &$ -18.41 \pm 0.1 $ &$<0.65$&$<154$ \\
18804 & PRIMER-UDS & 34.363515 & -5.176964 & - & - & $ 4.32 ^{+ 0.07 }_{- 0.27 }$& $ 1.5 \pm 0.1 $ & $ 45.1 \pm 0.4 $ &$ -19.21 \pm 0.1 $ &$<0.45$&$<114$ \\
18892 & PRIMER-UDS & 34.268908 & -5.176722 & - & - & $ 4.73 ^{+ 0.04 }_{- 0.20 }$& $ 1.0 \pm 0.1 $ & $ 44.4 \pm 0.2 $ &$ -17.34 \pm 0.1 $ &$<0.62$&$<123$ \\
19416 & PRIMER-UDS & 34.460763 & -5.175813 & - & - & $ 5.43 ^{+ 0.04 }_{- 0.06 }$& $ 1.9 \pm 0.1 $ & $ 44.8 \pm 0.3 $ &$ -18.32 \pm 0.1 $ &$<1.07$&$<221$ \\
21108 & PRIMER-UDS & 34.362963 & -5.173200 & - & - & $ 7.14 ^{+ 1.8 }_{-0.10 }$& $ 2.6 \pm 0.3 $ & $ 45.8 \pm 0.2 $ &$ -18.09 \pm 0.3 $ &$<0.62$&$<105$ \\
22773 & PRIMER-UDS & 34.438986 & -5.170543 & - & - & $ 4.40 ^{+ 0.21 }_{- 0.10 }$& $ 1.3 \pm 0.2 $ & $ 44.7 \pm 0.2 $ &$ -18.88 \pm 0.1 $ &$<0.57$&$<129$ \\
23575 & PRIMER-UDS & 34.408020 & -5.169147 & - & - & $ 5.66 ^{+ 0.02 }_{- 0.01 }$& $ 2.0 \pm 0.2 $ & $ 45.1 \pm 0.3 $ &$ -18.20 \pm 0.1 $ &$<0.76$&$<137$ \\
24081 & PRIMER-UDS & 34.346207 & -5.168197 & - & - & $ 5.66 ^{+ 0.02 }_{- 0.02 }$& $ 1.8 \pm 0.2 $ & $ 45.1 \pm 0.3 $ &$ -18.42 \pm 0.1 $ &$<0.52$&$<100$ \\
\enddata
\tablecomments{Sizes are measured in F444W band on the 0\farcs{04} images. The FWHM of the F444W PSF is 3.45 pixels.}
\end{deluxetable*}

\clearpage

\bibliographystyle{aasjournal}
\bibliography{refs}

\end{document}